\documentclass[useAMS,12pt]{article}

\usepackage{graphicx,epsfig,floatfig,epsf,scicite}

\usepackage{times}

\def\ltsim{\raise 2pt \hbox {$<$} \kern-1.1em \lower 4pt \hbox {$\sim$}}
\def\ltapprox{\raise 2pt \hbox {$<$} \kern-1.1em \lower 5pt \hbox {$\approx$}}
\def\gtsim{\raise 2pt \hbox {$>$} \kern-1.1em \lower 4pt \hbox {$\sim$}}
\def\gtapprox{\raise 2pt \hbox {$>$} \kern-1.1em \lower 5pt \hbox {$\approx$}}

\def\clust{Abell~3376 \,}
\def\com#1{$^\dagger$}
\def\deg{$^{\rm o}$}
\def\kms{km~s$^{-1}$}

\newenvironment{sciabstract}{%
\begin{quote} \bf}
{\end{quote}}

\newcounter{lastnote}
\newenvironment{scilastnote}{%
\setcounter{lastnote}{\value{enumiv}}%
\addtocounter{lastnote}{+1}%
\begin{list}%
{\arabic{lastnote}.}
{\setlength{\leftmargin}{.22in}}
{\setlength{\labelsep}{.5em}}}
{\end{list}}

\title{Giant Ringlike  Radio Structures  Around
Galaxy Cluster Abell 3376}

\author
{Joydeep Bagchi,$^{1\ast}$ Florence Durret,$^{2}$ Gast\~ao B. Lima Neto,$^{3}$\\ 
Surajit Paul$^{4}$ \\
\\
\normalsize{$^{1}$The Inter-University Centre for Astronomy and Astrophysics (IUCAA),}\\
\normalsize{
Post Bag 4,
Pune University Campus,
Pune - 411 007,
India}\\
\\
\normalsize{$^{2}$Institut d'Astrophysique de Paris, UMR-7095, CNRS,}\\
\normalsize{
Universit\'e Pierre
\& Marie Curie, Paris, France,}\\ 
\normalsize{and
Observatoire de Paris-Meudon, LERMA, Paris, France.}\\
\\
\normalsize{$^{3}$Instituto de Astronomia, Geof\'{\i}sica e Ci\^encias Atmosf\'ericas,}\\
\normalsize{
Universidade de S\~ao Paulo, S\~ao Paulo, Brazil - 05508.
}\\
\\
\normalsize{$^{4}$Institut fur Theoretische Physik und Astrophysik, Universitat Wuerzburg, }\\
\normalsize{
Sanderring 2, 97070 Wuerzburg, Germany.
}\\
\\
\normalsize{$^\ast$To whom correspondence should be addressed; E-mail: joydeep@iucaa.ernet.in}
}

\date{}

\begin{document} 




\maketitle 


\begin{sciabstract}
In the current paradigm of  cold
dark matter cosmology, large-scale structures are assembling through
hierarchical clustering of matter. In this  process, 
an important
role is played by  megaparsec (Mpc)-scale cosmic shock waves,  
arising in gravity-driven supersonic
flows of intergalactic matter onto  
dark matter-dominated collapsing  structures 
such as pancakes, filaments and clusters of galaxies. Here we report 
 Very~Large~Array telescope observations of   
 giant ($\sim 2$~Mpc~$\times \, 1.6$~Mpc), 
ring-shaped  non-thermal radio 
emitting structures, found at the outskirts of the rich cluster 
of galaxies Abell~3376.
These structures may trace  the 
elusive shock waves of cosmological large scale matter 
flows, which are   
energetic enough to 
power them. These radio sources may also be the 
acceleration sites
where magnetic shocks are possibly boosting 
cosmic-ray particles with energies of up to $10^{18}$ to $10^{19}$ 
electron volts.
\end{sciabstract}


{
A large fraction
($\sim 30 \%$) of the 
baryon mass of the universe at the present epoch 
resides in a  tenuous gas phase known as the 
warm-hot intergalactic medium (WHIM) \cite{Cen1999}. 
Although the WHIM is heated to 
$\approx 10^{5}$ to $10^{7}$ K by intergalactic shocks,  its direct detection is very
challenging owing to its low density and temperature. The main tracers of this 
gas are highly excited oxygen lines
mainly visible in soft X-rays and the far ultraviolet, whose emission or absorption signal 
is well below the sensitivity threshold of the current instruments.
Alternatively, the supersonic
infalls and resulting shock waves propagating in the 
magnetized inter-galactic medium (IGM) or WHIM  around galaxy clusters and filaments 
can be probed  by the synchrotron and inverse-Compton radiation emitted by 
energetic  electrons accelerated at the shock fronts 
\cite{Bagchi2002,Keshet2004}. For particles 
to be shock accelerated, 
magnetic fields need to be
present at the shocks for generation of Alfven waves (transverse
waves moving along magnetic field lines embedded in 
an electrically conducting fluid), and to-and-fro
scattering of particles across the shock front 
by wave-particle interactions [the so called 
diffusive shock acceleration process \cite{Drury83}]. 
The origin of cosmic magnetic fields 
is currently unknown \cite{Carilli2002}, and so
any observational evidence for them  in the IGM
environment is of great importance.
Magnetic fields ($\sim \mu$G) are observed in
the denser, and hotter (relative to WHIM)  
intra-cluster medium (ICM) of clusters of galaxies 
by means of the Faraday effect, causing rotation of the plane 
of polarization of light of the 
background radio sources \cite{2001ApJ...547L.111C}. Magnetic fields 
are further revealed by
the presence of cluster-wide Mpc scale (1 Mpc = $3.08 \times 10^{24}$ cm) 
radio-halos and peripheral 
radio-relics \cite{Bagchi_mnras98,Giovannini2004} (both  
are large-scale diffuse
radio sources, having no optical counterparts and no obvious connection
to the galaxies in clusters) in 
a few clusters of galaxies. Radio-halos and radio-relics are 
believed to be the result of synchrotron emission of relativistic
electrons shock accelerated in cluster mergers \cite{Ensslin98}. 

We have discovered with the Very Large Array (VLA) telescope at 1.4 GHz frequency,
a large-scale [$\sim 2 h^{-1} $ Mpc \cite{cosmo}]
ringlike synchrotron radio emission structure, possibly tracing the
intergalactic shocks  
around a rich cluster of galaxies \clust at
redshift z=0.046 \cite{Smith2004}.  
The southern galaxy cluster Abell~3376  was 
observed with the VLA,
a Y-shaped interferometric array of 30 25-meter-diameter radio
telescopes located on the plains of San Agustin in Socorro, 
New Mexico, the United States \cite{xrayanalys}. Figure~1 shows the most important 
findings;
a pair of giant, ``arc''
shaped diffuse radio-emitting sources, each with a linear dimension
$\sim 1 h^{-1}$ Mpc, located at the 
outskirts of this cluster at the
projected distance of $\sim 1 h^{-1}$ Mpc  from the cluster center.
In addition, the eastern structure shows several more 
thin filamentary structures behind the radio arc,
and a peculiar loop-like feature joining two linear filaments (Fig.~1).
This radio image is the result of combining 
data in the VLA CnB and DnC hybrid configurations, having the sensitivity for
mapping diffuse, large-scale emissions, as well as the 
angular resolution ($\sim$20  arc sec) for
identifying the superposed background point sources \cite{xrayanalys}.  
The VLA radio image shown
has $\sim 40 \, \mu$Jy per beam of root mean square noise background, 
and the  estimated signal-to-noise ratio
for detection of these giant radio arcs ranges from $\sim 3\sigma$ at the
faintest contour to $\sim 25\sigma$ at the peaks (after removing a
few superposed background point sources, which are visible in Fig.~1). Noticeably,
both radio structures are positioned with their concave side facing
the cluster center, and,- they fit quite well on the surface of a 
large, projected elliptical ringlike formation of dimension 
$\sim 2$~Mpc~$\times \, 1.6$~Mpc, oriented
in position angle $\approx 85^\circ$. The center of
this ellipse falls at the position  R.A. $\approx \,
06^h~01^m~32^s$, and Dec. $\approx \, -39^\circ 59^\prime
50^{\prime\prime}$, which is taken as the
center of symmetry of this cluster (marked `+` in Fig.~1). No other "radio-halo" 
type of diffuse, large scale 
emission was found near the
cluster center down to the surface brightness limit  mentioned above.

The luminous thermal bremsstrahlung
X-ray emission detected by the Position Sensitive Proportional Counter 
 (PSPC) detector 
onboard the Roentgen Satallite (ROSAT)  is
shown in Fig.~1 (in the 0.14 to 2.0~keV band). In the central region, 
it reveals a highly
disturbed, non-equilibrium state of the intra-cluster thermal gas,
which is obviously extended like a ``comet" or ``bullet-head", its wake  
extending along the major axis of the elliptical radio structure described
above. Both radio sources are located on the outer periphery of the 
X-ray emission observed by ROSAT, which has an unblocked field of view of
17.5 arcmin radius.  From galaxy redshift measurements 
in Abell~3376, a filamentary structure composed of at least three
major galaxy groups is found, oriented along  the general elongation of 
X-ray emission mapped by ROSAT and XMM-Newton (XMM: X-ray Multi Mirror Mission). 
\cite{Escalera94,Flin2006}. 

XMM-Newton and ROSAT 
X-ray observations \cite{xrayanalys} shown in Fig.~2
reveal strong evidence for merger activity of
subclusters along this filamentary axis.  
The field of view  of XMM is a circle of radius 15 arcmin, and 
the image shown is in the 
0.3 to 8.0 keV band.
The XMM  map of Fig.~2 is quite similar to the  ROSAT image in
Fig. 1, but it shows more clearly the ``bullet-head" and multiple X-ray
peaks to the south south-west of it, each one probably associated with
a merging group.  The X-ray temperature map 
(Fig.~2) and  has typical errors 
of about 10 percent. This map reveals an overall 
temperature of about 5~keV ($5.8 \times 10^{7}$ K),
with several alternating hot and cold regions crossing the cluster,
divided by a prominent ``cold-arc" at about 3~keV, which originates
at the north edge and curves southward toward the east.  The most
plausible scenario is that a large group or a small cluster is falling
onto the main body of the cluster from the east-northeast, thus
creating hotter regions through a shock. Such a scenario would agree
with the second-brightest galaxy on the X-ray peak associated with the
strong radio source MRC~0600-399, of which both radio-jets are bent
backwards towards east-northeast, away from the ``bullet", 
suggesting an infall and ram pressure on radio jets from a gaseous 
``wind'' blowing from the west-southwest direction along
position angel$\sim 70^\circ$ (Fig. 1, detailed image in fig.~S1). The dominant 
elliptical galaxy in \clust is located
in the south-west subcluster, $\sim 18$ arc min ($\sim 1$ Mpc) 
from the X-ray peak (Fig.~2).
The highly asymmetrical metal distribution [mainly iron, about $30 \%$
errors \cite{xrayanalys}]
near the center also
suggests a violent and recent dynamical event, where the gas is not
well mixed and we observe patches of high 
and low metallicity (Fig.~2;  metallicity  profile in fig.S2).

What is the energy source that could possibly power such giant radio
structures ? Any electron acceleration process must be at least 
energetic enough to account
for $\nu I_{\nu} \approx 2.1 \times 10^{40}$ erg/s of radio emission
observed at the frequency $\nu = 1.4$ GHz (sum of radio luminosity of two sides).
First, there is no evidence that any optical galaxy is
obviously associated with the radio arcs (Fig.~1), and it is
unlikely that these arcs are the usual cluster radio galaxies.  On the 
other hand, these radio
structures could be generic to radio relics known to occur in
certain merging clusters \cite{Bagchi_mnras98,Giovannini2004}, although
such double relics are rare. Another very similar configuration of
Mpc-scale binary radio relic arcs is  found  in the well known 
merging galaxy
cluster Abell~3667 (\cite{Huub1997}). Second, 
initial acceleration at a central
active-galactic-nucleus  like point source and particle 
transport by diffusion across a
$\sim$Mpc scale is not possible. The radiative energy loss timescale
$t_{IC}$ for an electron of relativistic Lorentz factor 
$\gamma$ is $t_{IC} \approx
2.3 \times 10^{8} \, \left( {\gamma \over 10^{4}} \right)^{-1} \,
(1+z)^{-4} \, {\rm year}$ [assuming only inverse Compton scattering
(IC) on the ambient 2.7 K background and a weak magnetic field $B<<3 \
\mu$G], and the diffusion length within the IC cooling time: $
l_{diff} \approx (D_{B} \ t_{IC})^{1/2} = 11.36 \left( {B / \mu G}
\right)^{-1/2}$ pc $<< 1$ Mpc, the scale of the observed radio
arcs [here $D_{B}= 1.7 \times 10^{23}\left( {\gamma \over 10^{4}} \right) 
\left( {B / \mu G}\right)^{-1} {\rm cm^{2}/s}$, the Bohm 
diffusion coefficient, for scattering on
saturated field fluctuations]. The discrepancy between the Bohm
diffusion length-scale and the radio structure size is so large that,
even with the inclusion of advective transport by bulk flows and more
effective diffusion in ordered magnetic fields, electrons are still
unable to cross the emission region within a radiative life-time.
Electrons must be accelerated to $\gamma \approx 1.8 \times 10^{4}$
($\sim 10$ GeV) for  synchrotron emission at frequency 
$\nu_{syn}=1.4$ GHz in a magnetic
field B=1$\ \mu$~G ($\nu_{syn} \propto \gamma^{2} B$). Sparse observations
\cite{Bagchi2002} and simulations \cite{Bruggen2005}
of  magnetic fields in the diffuse IGM at the cluster outskirts and
within filaments  suggest  that $B \sim 10^{-7}$ to $10^{-9}$ G. 
In this situation
the synchrotron loss timescale is $t_{syn} \approx 10^{3} ({B / 0.1 \mu
G})^{-2} (1+z)^{4} t_{IC} \ $, and IC radiation is the dominant cooling
process.  

Clearly, the detection of these large radio structures at $\sim$Mpc
distances from the cluster center inevitably requires some form of 
in situ acceleration mechanism for particles and the magnetic field
powering them. The detected structures cannot result from the chance
superposition of background radio point sources which are very few ($\sim 0.02$
sources per arcmin$^{2}$ above 1 mJy flux density),
and most of the emission is truly diffuse.  Lastly, it is very
unlikely that these diffuse sources are the radio-lobes of a 
currently active giant
radio galaxy (GRG), because to $\approx 0.2$ mJy/beam surface brightness
limit ($\approx 5 \sigma$ signal) no radio jets or plumes connect the
radio sources with any central optical galaxy, and GRGs of such 
extreme size ($\sim 2 $ Mpc) are not commonly found \cite{Schoenmakers01}.  
On the 
other hand, their concave bow-shock
like structure, symmetric and tangential juxtaposition on the merger
axis - tangential both to the chain of sub-clusters of galaxies and to
the X-ray emission elongation axis (Fig. 1) - strongly supports 
association of these arcs with the cluster,- and suggests their origin
in a cosmological-scale energetic event linked to the cluster
formation process.

Hydrodynamical simulations show that only shock waves induced in
structure formation processes are sufficiently extended, long-lived,
and energetic to overcome the strong radiation losses of the
relativistic electrons over a Mpc-scale, rapidly accelerating them to
relativistic energies \cite{Miniati2003,Keshet2003}, which could emit
synchrotron radio emission at the level observed \cite{Keshet2004}.  
These ubiquitous shocks are an inevitable consequence of 
gravitational collapse, and they are
pivotal to virialization of diffuse IGM.
Two competing shockwave inducing 
mechanisms are plausible:  
(i) outgoing 
``merger shocks'' emanating from the
cluster center, induced by  mergers of subclusters; and (ii) the 
accretion flows of IGM and
associated ``accretion shocks'' near the virial radius.
We briefly mention both these plausible models because their
observational signatures - which cannot be
discerned in the present data - are very similar. 

In the merger shock model,
a pair of outgoing, merger-generated shock waves \cite{Ricker2001} 
could have created these diffuse
radio sources on the outskirts of \clust by compressing
and accelerating a pre-existing lower energy ``fossil'' 
electron population \cite{Hoeft2004}.  
It takes $\sim 10^{9}$  years
for shock fronts to cross the $\sim 1$-Mpc distance from the cluster center 
(at the `+' mark in Fig.~1) if the shock  
speed is $V_{s} = 10^{3}$ km/s. The merger shock  model adequately explains the extended
steep-spectrum radio arcs and elongated X-ray structure found in 
a well-studied 
cluster Abell~3667 \cite{Roettiger99}. This is a plausible model given the 
striking  resemblance of  
\clust and Abell~3667, both of which provide strong evidences for mergers and 
have similar X-ray and radio morphologies \cite{Huub1997}. The peripheral
location of quasi-ring-like radio arcs and the absence of a central radio halo
in \clust agree with  simulations of merger 
shocks [\cite{Hoeft2004};
and fig.~S3].

In the  accretion model, these giant radio structures could be tracing 
the accretion
shocks from infall of IGM at the outskirts of the cluster Abell~3376. 
An accretion shock at the virial radius propagates
outward while infalling gas crosses it inward supersonically, which
could lead
to shock acceleration of particles and radio emission 
if the infalling gas is magnetized [as in \cite{Ensslin98}].
The virial radius is $r_{vir}={[3 \ M/(4
\pi \Delta_{c}(z) \rho_{crit}(z))]^{1/3}}$, where $\Delta_{c}(z) $ is
the ratio of mass density to critical density $\rho_{crit}(z)$ \cite{cosmo} inside
a dissipationless, spherical, collapsing virializing halo of mass M at
redshift z. In Λ$\Lambda$ Cold-Dark-Matter
and Einstein-de Sitter  (\cite{cosmo})
cosmologies $\Delta_{c} \approx 340$ and $\Delta_{c}
\approx 178$ respectively, at $z=0$ \cite{Kitayama96}, 
implying that $r_{vir}=$ 1.4 to 1.7
Mpc for Abell~3376,- and the cluster virial mass is
$M_{vir} = 5.2 \times 10^{14}$ solar masses
($M_{\odot}$) \cite{Girardi1998}.  This
exceeds the observed $\sim 1$-Mpc distance at which radio sources are
found, possibly due to a  non spherical filamentary 
geometry  (Fig.~1) and an apparent
linear foreshortening due to  projection on the observer's sky plane.

Recent hydrodynamical simulations reveal an 
intergalactic shock  structure more complex than assumed in  
a simple spherical accretion model described above.
Around rich clusters
these shocks (both accretion and merger) do not form 
fully illuminated rings (in projection), but
they are strongest along the axes where filaments funnel the IGM deep into
the forming clusters \cite{Miniati2003,Keshet2003}. 
Along the same directions, high-speed 
mergers of subclusters (with merger shocks) may take place, as observed 
here  and in cluster Abell~3667 \cite{Huub1997}.
This may provide a clue to the origin of the radio structures
observed in Abell~3376, which are located on an elliptical
ring-like formation in projection (Fig.~1), yet only sections of this ring are
actually illuminated along an axis, where X-ray data  clearly show that 
mergers and infalls are
taking place and where shocks are strongest, leading to more efficient
particle acceleration. The actual three dimensional morphology
could be like a ``bubble'' or ``egg-shell'' on the surface of which the
shocks are located within a narrow zone. Some
recent numerical simulations of structure 
shocks around similar massive galaxy
clusters  ($10^{14}$ to $10^{15} \, M_{\odot}$) do predict  this 
quasi ring-like geometry of non thermal 
radiation \cite{Keshet2003,Hoeft2004}.

Therefore, in the giant radio
structures discovered in \clust (and Abell~3667), 
we are probably witnessing the first observational evidence for
merger or accretion (or even both) shocks near the sparsely studied 
virial infall region of a 
massive galaxy
cluster. These shocks are an important  probe of the gas dynamics 
at the transition zone between the hot cluster medium and cold WHIM gas.  
The Mach numbers ($M_{s}$) associated with these shock structures are
not known as yet. Their estimation requires temperature data on both
sides of the shock to fit Rankine-Hugoniot equations. For an
intergalactic  
shock at the  cluster outskirts in cold medium, simulations predict 
high Mach numbers [$M_{s}>>1$ \cite{Gabici2003}], and the pre and 
post shock temperatures $T_{1}$, and $T_{2} $ are
related as $T_{2}/T_{1}\approx (5/16)M_{s}^{2} $ in a gas of specific
heat ratio $\Gamma = 5/3$. Setting $T_{2} \sim T_{vir} = 4 \times 10^{7}$
K, the virial temperature ($T_{vir}$) for the  virial mass of
the cluster, the temperature of the infalling  gas
spans the range $\sim 5 \times 10^{4-6}$~K, for  
$M_{s}$ between 50 and 5. This is close to the expected temperature range
of WHIM gas \cite{Cen1999}, 
although too cold to be
detectable in a  keV-band X-ray observation (Fig.~1).

The exact physical mechanism of the electron
acceleration is yet to be revealed, but diffusive shock acceleration (DSA) is
a strong possibility, which explains the $\sim 4$ Mpc scale diffuse radio
emission from the filamentary  proto-cluster ZwCl~2341.1+0000 \cite{Bagchi2002}. 
In DSA, the time scale for a particle (electron, proton, or a heavier ion) 
to reach energy E is \cite{Drury83};
$t_{acc}(E) \approx 8 D_{B}/{u_{3}^{2}} = 8.45 \times 10^9 \, u_{3}^{-2} E_{19}
B_{\mu}^{-1} Z^{-1}$  year (for a strong shock of $M_{s}>>1$, and Bohm diffusion).
Here, $E_{19}$=$(E/10^{19} \, eV)$,
$u_{3}$=$(V_{shock}/10^{3}$ \kms), $B_{\mu}$=$(B/10^{-6} \, G)$, and $Z$
is the ionic charge ($Z = 1$ for electron or proton).
For cosmic-ray
protons, which suffer negligible radiative losses below 50 EeV, the
highest acceleration energy, $E_{max}^{P}$, is limited only by the
finite lifetime of shocks, i.e., $t_{acc}=t_{merger}\sim 10^{9}$ to $10^{10}$ years 
(here,
$t_{merger}$ is the time scale for merger),
thus giving $E_{max}^{P} \sim 10^{18-19}$ eV. Heavier ions 
carrying more charge 
(for example, $^{26}$Fe) could be accelerated up to much larger energy,
limited by loss processes.  For cosmic-ray electrons, substantial
radiative losses would limit their energy to a maximum $E_{max}^{e} \sim
3.73 \times 10^{13} \, u_{3} \, B_{\mu}^{1/2}$ {\rm eV}. Lower-energy
electrons of $\sim 10 \, B_{\mu}^{-1/2}$ GeV 
would radiate the observed 1.4 GHz synchrotron
radio emission.  The sketched scenario is consistent with 
hydrodynamical simulations \cite{KJ2005}, which show that
during gravitational infall and violent mergers in clusters, protons can be
accelerated up to the  energy $E_{max}^{P} \sim 5 \times 10^{19} $ eV,
if a turbulent magnetic field of $\sim$ 0.1 to 1 $\mu G$ is available, and if
a fraction about $10^{-4}$ of the infalling kinetic energy can be
injected into intergalactic medium  as high-energy particles. Our estimate of 
the minimum energy magnetic
field ($B_{min}$) within the radio-arc regions in \clust is 
 $B_{min} = 0.5$  to 3.0 $\mu$G, which
depends on model parameters \cite{B_equipart}.
 
Our observations here represent a substantial advance in the field, because they 
probe several important components of the
cosmic environment: intergalactic gas, magnetic fields and
cosmic-rays.  They indicate that magnetic fields of appreciable
strength are present not only in the ICM but also in the diffuse
inter galactic medium, i.e., in the gas that will be shocked as it
accretes onto collapsing structures - the precursors of virialized
galaxy clusters. These magnetic fields are necessary 
for providing the scattering centers for the diffusive shock
acceleration mechanism, and also for the synchrotron
emission that we have observed.  Because it is not obvious how
magnetic fields are amplified up to such large values along filaments,
our observation poses further challenges to theoretical models
\cite{Kulsrud97,kronberg99}.

On the other hand, as we show in this work, if diffusive shock
acceleration takes place with some efficiency during the non-linear
stage of large-scale structure formation, cosmic-ray ions accumulating
in the forming structure could become dynamically important \cite{mrkj01}.  
Objects like \clust could  be 
acceleration sites for
cosmic-rays, where magnetized shocks 
are possibly boosting hadronic
particles to
relativistic energies up to $10^{18}$ to $10^{19}$ eV. Downstream
from the shocks (toward the cluster center), accelerated
protons will  be transported by diffusion and advection, 
over a length scale ($l_{diff}$) 
comparable to the cluster size,
$l_{diff} \approx 1.08 \, u_{3}^{-1} E_{19} B_{\mu}^{-1}$ Mpc, under
the Bohm diffusion regime \cite{Gabici2003}. These protons will remain trapped within
the cluster volume for a time scale comparable to the age of the
cluster until their energy approaches $E^{P} \, \gtsim \,\, 2 \times
10^{17}\ B_{\mu}$ eV.
Inverse Compton scatter of the 2.7 K cosmic microwave background
photons from both primary electrons accelerated in shocks and from
secondary electrons  originating in hadronic processes 
(mainly pion decay and pair production), 
will give 
rise to photons of energies $E_{\gamma}
\sim 100~{\gamma^{e}_{7}}^{2}$ GeV (where ${\gamma^{e}_{7}}$ is the
electron Lorentz factor in units of $10^7$). 
This inverse Compton spectrum
could extend upto $\sim 10 $ TeV, limited by  a likely high-energy
cut-off in
the electron energy spectrum. Radiation from primary electrons would
trace the current shock locations, owing to their 
short radiative cooling time.
This tell-tale
radiation might be detectable with hard
X-ray (SUZAKU, NEXT), or 1 MeV to 100 GeV $\gamma$-ray 
(GLAST), or high-energy TeV $\gamma$-ray (CANGAROO, MAGIC, VERITAS, HESS)
telescopes, opening a new window on the non-thermal processes in 
cosmological large-scale flows.

}

\bibliography{Science_draft}

\begin{thebibliography}{10}

\bibitem{Cen1999}
R.~{Cen}, J.~P. {Ostriker}, {\it Astrophys. J. \/} {\bf 519}, L109 (1999).


\bibitem{Bagchi2002}
J.~{Bagchi}, {\it et~al.\/}, {\it New Astronomy \/}, {\bf 7},
Issue 5, 249 (2002).


\bibitem{Keshet2004}
U.~{Keshet}, E.~{Waxman}, A.~{Loeb}, {\it Astrophys. J. \/}
{\bf 617}, 281 (2004).


\bibitem{Drury83}
L.O'C. {Drury}, {\it Rep. Prog. Phys. \/} {\bf 46}, 973 (1983).


\bibitem{Carilli2002}
C.~L. {Carilli}, G.~B. {Taylor}, {\it Ann. Rev. Astron. and Astroph. \/} {\bf 40}, 319 (2002).

\bibitem{2001ApJ...547L.111C}
T.~E {Clarke}, P.~P. {Kronberg}, H.~{B{\" o}hringer}, {\it Astrophys. J. \/} 
{\bf 547}, L111 (2001).

\bibitem{Bagchi_mnras98}
J.~{Bagchi}, V.~{Pislar}, G.~B. {Lima Neto}, 
{\it Mon. Not. R. Astron. Soc. \/} {\bf 296}, L23 (1998).

\bibitem{Giovannini2004}
G.~{Giovannini}, L.~{Feretti},
{\it J. of the Korean Astron. Soc. \/} {\bf 37}, 323 (2004).


\bibitem{Ensslin98}
T.~A {En{\ss}lin}, P.~L. {Biermann}, U. {Klein}, S.~{Kohle}, 
{\it A \& A \/} {\bf 332}, 395 (1998).

\bibitem{cosmo}
{\footnotesize We have
used a Hubble constant of $H_{0}=$ h 70~km/(s\-\ Mpc), where h is 
a dimensionless 
scaling parameter. We adopt a ``concordance'' model of big bang cosmology: the 
$\Lambda$-CDM (lambda cold dark matter) cosmological model 
defined by matter (dark matter + baryonic matter) 
density $\Omega_{M}=0.3$, and dark energy 
(denoted by cosmological constant $\Lambda$)
density $\Omega_{\Lambda}=0.7$. Thus, $\Omega_{M}+\Omega_{\Lambda}=1$. 
The Einstein-de Sitter cosmology has no dark energy, i.e. $\Omega_{\Lambda}=0$,
but $\Omega_{M} =1$. All
the dimensionless densities ($\Omega_{M}, \Omega_{\Lambda}$) are defined 
relative to the critical density
for closure of the universe, $\rho_{crit} = (3 H_{0}^{2})/(8 \pi G)$, 
where G is
the gravitational constant.  The $\Lambda$-CDM cosmology
results in 0.904 kpc per arcsec plate-scale
for redshift $z = 0.046$.}

\bibitem{Smith2004}
R.J.~{Smith}, {\it et~al.\/}, {\it Astron. J.  \/} {\bf 128}, 1158 (2004).


\bibitem{xrayanalys}
 ``Materials and methods'' are available as supporting material on 
``Science Online''. 


\bibitem{Escalera94}
E.~{Escalera}, {\it et~al.\/}, {\it Astrophys. J. \/}
{\bf 423}, 539 (1994).


\bibitem{Flin2006}
P.~{Flin}, J.~{Krywult}, {\it Astron. Astrophys. \/} {\bf 450}, 9 (2006).


\bibitem{Huub1997}
H.~{Rottgering}, M.H.~{Wieringa}, R.W.~{Hunstead}, R.D.~{Eckers}, 
{\it Mon. Not. R. Astron. Soc. \/} {\bf 290}, 577 (1997).





\bibitem{Bruggen2005}
M.~{Bruggen}, {\it et~al.\/}, {\it Astrophys. J. \/} {\bf 631},  L21-L24 (2005).

\bibitem{Schoenmakers01}
A.P.~{Schoenmakers}, A.G.~{de Bruyn}, H. J. A.~{Rottgering}, 
H.~{van der Laan},  
{\it  Astron. Astrophys.  \/} {\bf 374}, 861 (2001).



\bibitem{Miniati2003}
F.~{Miniati}, {\it Mon. Not. R. Astron. Soc. \/} {\bf 342}, 1009 (2003).

\bibitem{Keshet2003}
U.~{Keshet}, E.~{Waxman}, A.~{Loeb}, V.~{Springel}, 
L.~{Hernquist}, {\it Astrophys. J. \/} {\bf 585}, 128 (2003).


\bibitem{Ricker2001}
P.M.~{Ricker}, C.L.~{Sarazin}, {\it Astrophys. J. \/} {\bf 561}, 
621 (2001).




\bibitem{Hoeft2004}
M.~{Hoeft}, M.~{Bruggen} \& G.~{Yepes}, {\it Mon. Not. R. Astron. Soc. \/}
{\bf 347}, 389 (2004).

\bibitem{Roettiger99}
K.~{Roettiger}, J.~O. {Burns}, J.~M. {Stone}, {\it Astrophys. J. \/}
{\bf 518}, 603 (1999).


\bibitem{Kitayama96}
T.~{Kitayama}, S.~{Yasushi}, {\it Astrophys. J. \/} {\bf 469}, 480 (1996).


\bibitem{Girardi1998}
M.~{Girardi}, G.~{Giuliano}, F.~{Mardirossian}, 
M.~{Marino}, W.~{Boschin}, 
{\it Astrophys. J. \/} {\bf 505}, 74 (1998).



\bibitem{Gabici2003}
S.~{Gabici}, P.~{Blasi}, {\it Astrophys. J. \/} {\bf 583}, 695 (2003).

\bibitem{KJ2005}
H.~{Kang}, T.W.~{Jones}, {\it Astrophys. J. \/} {\bf 620}, 44 (2005).


\bibitem{B_equipart}
{\footnotesize
The minimum energy magnetic field $B_{min}$ is obtained
by minimizing the total nonthermal energy density ($u_{min}$), which consists of
the magnetic field energy and the energy contained in all relativistic 
particles. Specifically, $B_{min} \propto {(1+k)^{2/7}}{{(\phi \, V)}^{-2/7}}
{L_{syn}^{2/7}}$. In this expression, the  
model parameters are as follows: $k$, the ratio of energy in
heavy particles to electrons; $\phi$, the fraction of source volume 
filled by magnetic fields; and V, the total source volume. The 
total synchrotron 
radio luminosity, $L_{syn}$, can be obtained from observations 
by integrating the radio
spectrum between the lowest and highest frequencies 
(we used 10 MHz to 100 GHz). Parameters K, $\phi$, and V are usually unknown 
and need to be guessed reasonably. For our estimate of $B_{min}$, we
used the values within range: $K =$ 1 to 100, $\phi =$ 0.1 to 1, line-of-sight
depth = 270 kpc (for calculating V), and spectral index $\alpha =0.5$
(for obtaining $L_{syn}$).
For further details, see (31)}.  

\bibitem{Kulsrud97}
R.~{Kulsrud}, R.~{Cen}, J.P.~{Ostriker}, \& D.~{Ryu}, 
{\it Astrophys. J. \/} {\bf 480}, 481 (1997). 




\bibitem{kronberg99}
Kronberg, P.~P., Lesch, H. \& U.~{Hopp}, {\it Astrophys. J. \/} {\bf 511}, 56 (1999).


\bibitem{mrkj01}
F.~{Miniati}, D.~{Ryu}, H.~{Kang}, \& T.~W. {Jones}, 
{\it Astrophys. J. \/} {\bf 559}
, 59 (2001).

\bibitem{Govoni04}
F.~{Govoni}, L.~{Feretti},
{\it Int. J. Mod. Phys. (D) \/} {\bf 13}, no. 8, 1549 (2004).



\end{thebibliography}

\bibliographystyle{Science}


\begin{scilastnote}
\item F.D. acknowledges support from CNES and PNC, CNRS/INSU. G.B.L.N. 
acknowledges support from CNPq, CAPES/Cofecub Brazilian-French 
collaboration, and FAPESP through the Thematic Project 01/07342-7. 
S.P. acknowledges
The Deutsche Forschungsgemeinschaft
(DFG), Research Training Groups (RTG) 1147 for financial support and specially 
thanks IUCAA for support. The National Radio Astronomy Observatory is 
a facility of the National Science Foundation operated under 
cooperative agreement by Associated Universities, Inc. The X-ray data is based on 
observations obtained with XMM-Newton, an ESA science mission with 
instruments and contributions directly funded by ESA Member States and NASA.
\end{scilastnote}
\noindent
{\bf Supporting Online Material} \hfill \break
www.sciencemag.org \hfill \break
SOM Text \hfill \break
Figures: S1 to S3 \hfill \break
References and Notes \hfill \break
\vskip 1cm
12 June 2006; accepted 06 October 2006 \hfill \break

\clearpage
\begin{figure}
\hskip 2cm {\bf (A)}
\begin{center}
\includegraphics[width=7.75cm ,angle=-90]{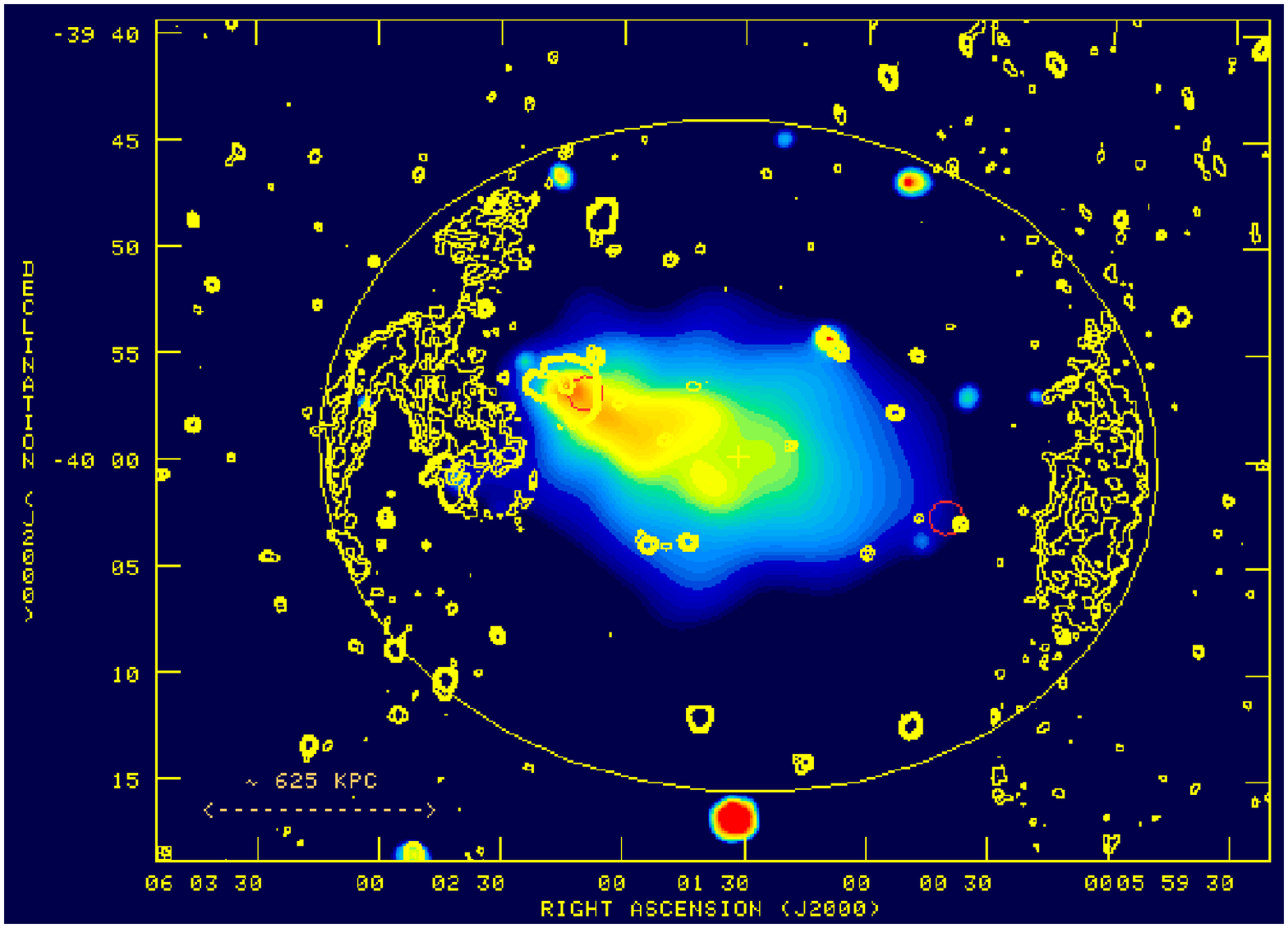}
\end{center}
\hskip 2cm {\bf (B)}
\begin{center}
\includegraphics[width=5.99cm]{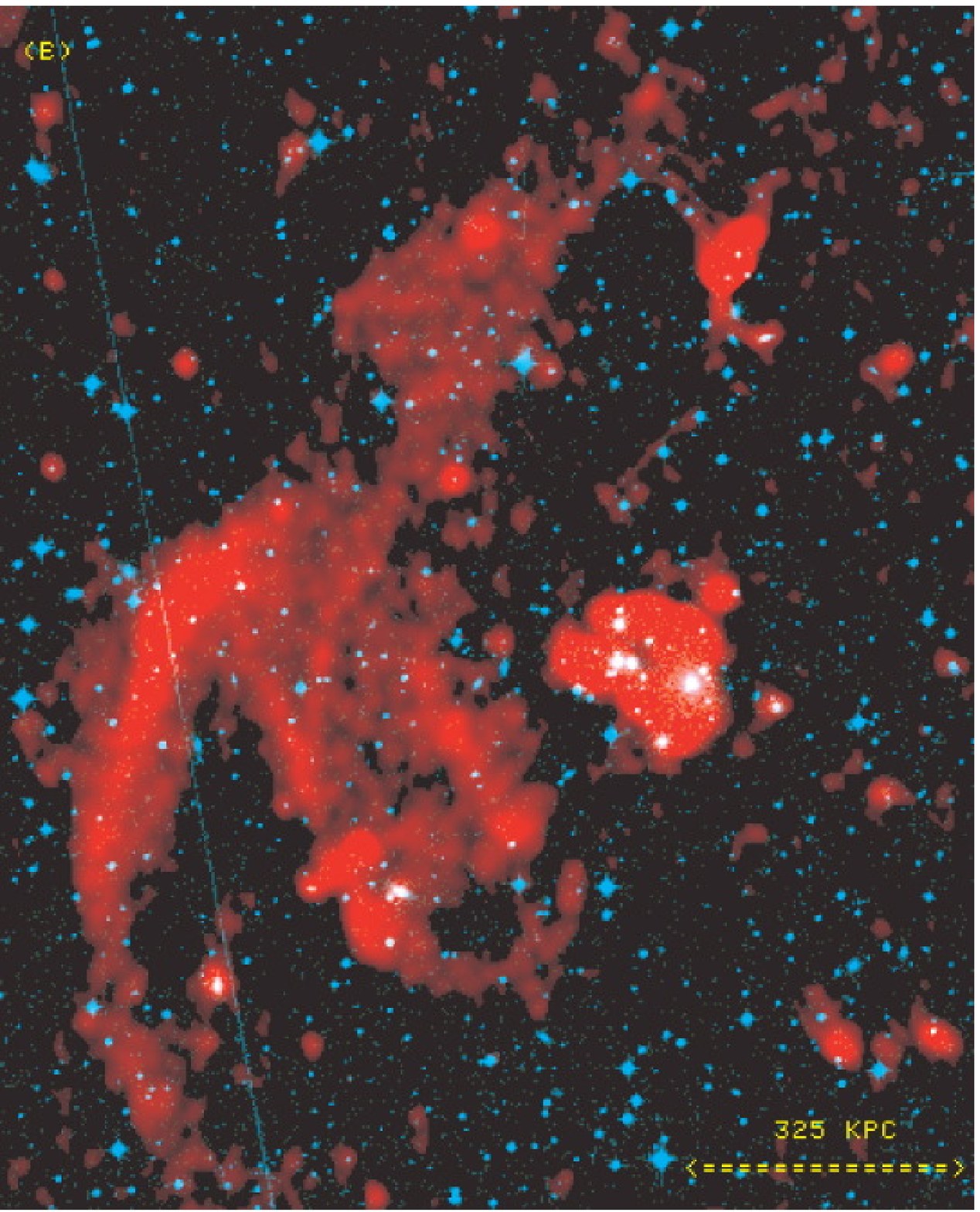}
\includegraphics[width=4.5cm]{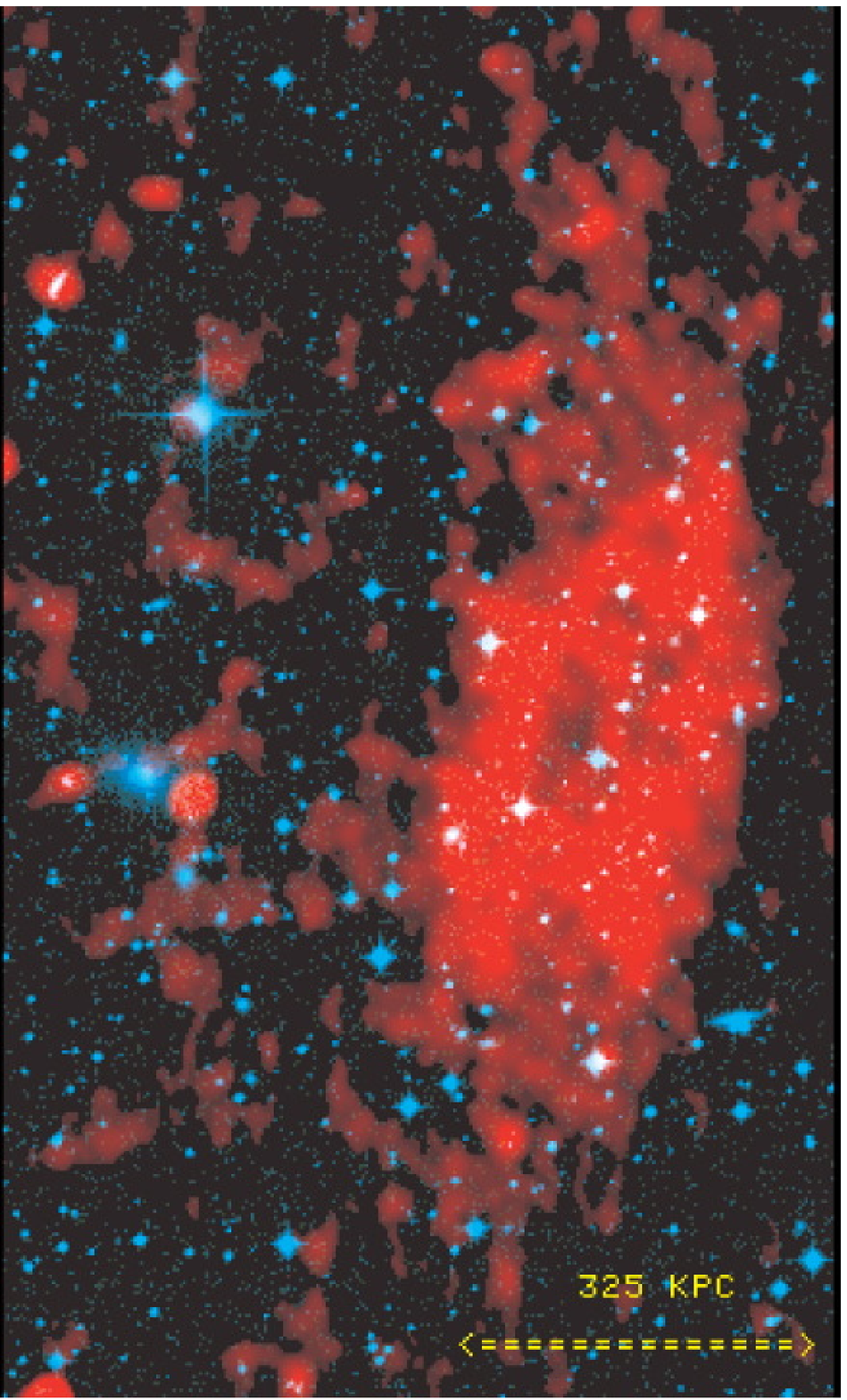}
\end{center}
\caption {\footnotesize {{\bf (A)} A composite map of radio and
X-ray emissions from the galaxy cluster Abell~3376.  The radio emission is
represented by yellow contours (0.12, 0.24, 0.48 and 1 mJy per beam. Beam
width: 20 arcsec FWHM Gaussian) obtained from the Very Large Array 1.4 GHz
observations \cite{xrayanalys}.  The yellow ellipse shows 
an  elliptical fit to the
peripheral radio structures, and the `+' marks the center of the
ellipse. The central color image depicts the thermal
bremsstrahlung X-ray emission detected by the Position Sensitive
Proportional Counter instrument
onboard the Roentgen Satellite ($\approx$12-ks exposure,  
within 0.14 to 2.0~keV band).  The red
circles mark the position of the two brightest cluster galaxies - the
brightest elliptical galaxy on the lower-right and the second brightest elliptical 
galaxy
associated with the bent-jet radio source MRC~0600-399 near the X-ray
peak.  {\bf (B)} Composite images obtained from  superposing the 
radio and optical
images.  The VLA 1.4 GHz radio maps (in red) for
the eastern (left) and the western (right) radio structures are shown
overlayed on the red band Digitized Sky Survey image (in blue). 
}}
\end{figure}

\clearpage
\begin{figure}
\begin{center}
\includegraphics[width=15cm]{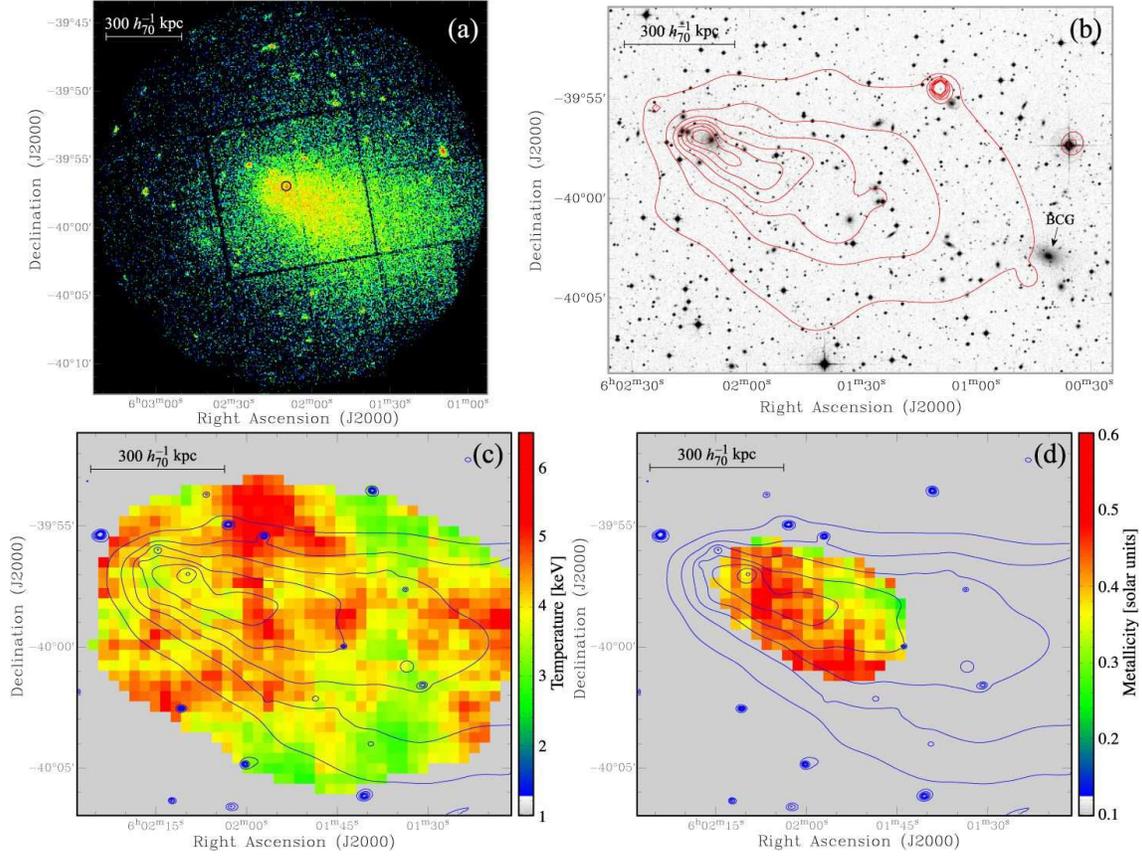}
\end{center}
\caption{\footnotesize {{\bf (a)} XMM-Newton MOS1 and MOS2 
[MOS stands for Metal Oxide Semiconductor \cite{xrayanalys}] telescope composite
photon image in the 0.3 to 8.0 keV band. The black circle shows the position
of the second brightest galaxy, which coincides with the X-ray peak.
{\bf (b)} Red band Digitized Sky Survey optical image with ROSAT 
observed smoothed X-ray intensity iso-contours superposed. The ROSAT energy range 
is 0.2 to 2.0 keV.
The brightest cluster
galaxy (BCG) is indicated with an arrow. The BCG sits at the edge of the
ROSAT field of view but outside of the XMM field of view.  {\bf (c)}
Temperature map (color scale in keV units) 
derived from XMM data. Noticeably, there are several alternating cold and
hot regions, their temperature varying from $\sim 2$ to $\sim 6\,$keV.
The superposed X-ray intensity contours are from an adaptively
smoothed XMM image.  {\bf (d)} Metallicity map (color scale in solar units) 
derived from XMM data \cite{xrayanalys}. It
shows a strong metallicity variation  along the X-ray intensity elongation axis. 
Contours are the same as in the previous
panel.  The scale bars are in kpc, assuming redshift $z = 0.046$ and standard
$\Lambda$ cold dark matter cosmology \cite{cosmo}.}}
\end{figure}

\clearpage

\section*{\Large Supporting Online Material}
\vskip 1.5cm
\subsection*{1. VLA observations and data analysis}

The southern galaxy cluster Abell~3376  was observed with the Very Large
Array (VLA) facility of National Radio Astronomy Observatory (NRAO), a Y-shaped 
interferometric array of 30 25m-diameter radio
telescopes located on the plains of San Agustin Socorro, New Mexico, USA
({\it 1,\ 2}). The L-band observations were done on two separate days: on 20th
September 2002, in ``CnB" configuration (6 hours total) and on 25th January
2003 in ``DnC" configuration (6 hours total). The ``CnB" and ``DnC" are hybrid
configurations in which the antennas on the east and west arms are moved in
for the shorter configuration (C or D), but those on the north arm remain in
extended confuguration (B or C). This results in better  
mapping of sources located
in the extreme southern sky such as Abell~3376, due to a resultant more 
circular synthesized beam shape.
The basic data produced by the array are the visibilities, or measures 
of the spatial coherence function, formed by correlation of signals from 
the array's elements (one visibility sample for each pair of antennas, 
351 visibilities
during every sampling from all telescope pairs).
The two IFs (intermediate frequencies)
were centered at 1464.9 MHz (IF1) and 1385.1 MHz (IF2) with bandwidths of 50
MHz in each IF. The data were obtained in radio continuum mode and amplitude
(flux density scale) calibration was obtained by boot-strapping to 3C~48 which
was assumed to have a flux density of 15.6170 Jy at IF1 and 16.3186 Jy at IF2
respectively. We used VLA calibrator 0616-349 as phase calibrator to track
visibility phases. As the angular width of the entire cluster is quite large ($>
40$ arcmin), therefore each observation contained 3 separate telescope
pointings: centered on the cluster center, the east radio relic and west radio
relic respectively. At 21cm observing wavelength, the VLA system temperature 
is $\sim$45 K, when pointed at low elevation, required for pointing at 
$-40$\deg \, source declination. The individual dish antennas have beam 
diameter (effective field of view) of $\sim$30 arcminutes at 1.4 GHz.

For analysis we used the NRAO `AIPS' package. Bad data points from faulty
antennae/baselines and from high amplitudes or corrupted phases were edited
out. Radio imaging of very faint, extended sources is always challenging due
to the large primary beam area of the antennae, resulting in detection of many
strong sources over this area. In addition, the effects of sky-curvature
cannot be neglected in mapping and cleaning of fields large in size. In order
to mitigate these problems as much as possible, for imaging and deconvolution
we used the task `IMAGR' that permits the `faceted' wide-field mapping and
cleaning option. We mapped up to 30 small fields (facets) centered on strong
sources located a priori within the field of view centered on each of the
three pointings, and then simultaneously cleaned the entire set. At the end of
the cleaning cycle the clean components from all facets together were restored
back onto a single field to obtain a large wide-field image. We point out that
due to the use of a separate tangent point for each field, 3-D imaging, and the
subtraction of clean components from un-gridded data (thus preventing 
aliasing of
side-lobes), the quality of the final clean image is quite good.

During
imaging, in order to emphasize the large-scale features and suppress the
oscillations in the side-lobes, the visibility data were tapered by a Gaussian
that had $30\%$ weight at the 30~k$\lambda$ length of the interferometric
spacing and parameter ROBUST = 5 was used, corresponding to `natural'
weighting of interferometric visibilities. The cleaned maps were restored with
circular Gaussian beams of $20^"$ FWHM width. The calibrated data from two
hybrid configurations were combined with the task `DBCON'. This has resulted
in better sensitivity and excellent mapping of diffuse large-scale features,
as well as angular resolution needed to identify the superposed background
point sources. We did at least four passes of phase and amplitude
self-calibration ({\it 3}) to correct for the
closure phase and amplitude errors. The
background noise on the final radio images (Fig.~1 and 2 main paper,
Fig.~S1 this section) 
reaches down to $\sim 30-40 \mu
\,$Jy/beam of RMS and the S/N ratio for detection of diffuse radio features
(shock fronts) ranges from $\sim 3 \sigma$ at the faintest contour to $\sim 25
\sigma$ at the peaks (after removing a few superposed background point
sources). We used the task `PBCOR' to correct for the primary beam attenuation
effect of individual telescopes and all flux density measurements were done on
these corrected images. The point source corrected flux densities for the east
and west radio arc regions are 166 mJy and 136 mJy respectively.
\subsection*{2. XMM-Newton and ROSAT observations and data analysis}

The ROSAT (Roentgen Satallite) PSPC-B (PSPC: Position Sensitive
Proportional Counter
detector) X-ray data were retrieved from the High Energy Astrophysics Science
Archive Research Center (HEASARC). The observation was performed on February
28, 1992 with a 11.7~ks exposure time. From the photon list file we have made animage in the [0.14--2.0]~keV band with 16 arcsec pixels (the spatial resolution
is 25 arcsec). This image is shown in Fig.~1 and Fig.~2 of the main paper.

More recent X-ray imaging and spectroscopy data was obtained with XMM-Newton
telescope, providing much higher resolution and sensitivity.
These data are shown in Fig.~2 of the main paper.
XMM-Newton is the most sensitive X-ray
telescope  built until now, thanks to its 58 mirrors (Wolter I
grazing-incidence mirrors which are nested in a coaxial
and cofocal configuration) collecting
more X-rays than any other previous X-ray observatory. XMM-Newton's
scientific objective is to perform spectroscopy of cosmic X-ray sources
over a range of energies from around 0.15 keV to 15 keV.
European Photon Imaging Camera (EPIC) onboard XMM-Newton  is the
main focal plane instrument providing CCD imaging and spectroscopy
at the focus of three XMM-Newton mirror systems.
EPIC consists of two MOS (Metal Oxide Semiconductor) CCD cameras, and
one PN-type CCD camera. Each MOS camera contains an array of
seven MOS CCDs; and the PN-camera contains an array of 12 PN-type CCDs.
The three EPIC cameras together offer the possibility to perform extremely
sensitive imaging observations over the telescope's field of
view (FOV) of 30 arcmin diameter, and in the energy range from 0.15 to 15
keV with moderate spectral ($E/∆\Delta E \sim 20-50$) and angular
resolution (point spread function, about 6 arcsec FWHM).

We retrieved from the XMM-Newton archive a 47.2~ks observation performed on
April 01, 2003 in standard Full Frame mode (Observation ID: 0151900101, Proposal
no.: 15190). The pointing center of the telescope was on the position
R.A. $
06^h~02^m~8.6^s$, and Dec. $ -39^\circ 57^\prime
18.0^{\prime\prime}$ (J2000).
The basic data processing (the
``pipeline'' removal of bad pixels, electronic noise and correction for charge
transfer losses) was done with the XMM Science Analysis Software (SAS) V6.0.
For the spectral analysis, we have used the observations made with the
EPIC/MOS1, MOS2 and PN cameras. For the EPIC/MOS cameras, after applying the
standard filtering, keeping only events with FLAG~$=0$ and PATTERN $\le$ 12,
we have removed the observation times with flares using the light-curve in the
[10--12 keV] band using a 3-sigma clipping technique. The cleaned MOS1 and MOS2
event files have remaining exposure times of 19.3~ks and 19.7~ks,
respectively. We used a similar procedure for the PN, applying the standard
filtering and keeping only events with FLAG~$=0$ and PATTERN $\le$ 4. The
cleaned PN event file has a useful exposure time of 19.2~s. The redistribution
and ancillary files (RMF and ARF) were created with the SAS tasks
\texttt{rmfgen} and \texttt{arfgen} for each camera and each region that we
analysed.

Spectra were analyzed with \textsc{xspec}~11.3.2. Since the spectra were
rebinned, we have used standard $\chi^{2}$ minimization. We have applied MEKAL
({\it 4,5}) plasma models. The metal abundances (or metallicities) are based
on the solar values given by ({\it 6}). We chose to use these abundances for
easier comparison with previous work, even though these values may not be
accurate - more recent solar abundance determinations, e.g. ({\it 7,8}), give
lower O and Fe abundances than ({\it 6}). Using the combined XMM-Newton data
the only clearly detected line is the Fe XXV K-$\alpha$ complex at $\sim
6.7$~keV (restframe energy), thus the metallicity is measured
based mainly on the
iron abundance. The photoelectric absorption -- mainly due to neutral hydrogen
in our galaxy -- was computed using the cross-sections given by ({\it 9}),
available in \textsc{xspec}.

In order to obtain the mean temperature and metallicity of the cluster, we
extracted the X-ray spectrum in an ellipse centered on the X-ray peak with a
semi-major axis of 440 arcsec and semi-minor axis of 390 arcsec. The spectral
fits were done with data in the range [0.3--8.0~keV], simultaneously with
MOS1, MOS2, and pn. Spectral fits were done on the  background corrected
data. We use the standard background obtained at the XMM science center,
created by adding sky exposures of intermediate galactic latitude without
bright sources ({\it 10}). Within the region selected for spectral fits,
the mean emission-weighted temperature is $kT = 4.1 \pm
0.3\,$keV and the mean metallicity is $Z = 0.3 \pm 0.1 Z_{\odot}$, where
$Z_{\odot}$ denotes solar metal abundance. The
bolometric X-ray luminosity in this region is $2 h_{70}^{-2}\times
10^{44}\,$erg~s$^{-1}$.

We have produced the thermal emission, temperature and
metallicity maps (Fig.~2) using an
adaptively kernel technique
described in ({\it 11}). Briefly, we do a pixel-by-pixel spectral fit, adapting
the size of the fitted region based on a minimum number of net counts. Each
fitted pixel is then assigned with a temperature or metallicity value. For
the emission map, the pixel size is 3.2 arcsec. For the temperature and metal
abundance maps, the pixel size is 12.8 arcsec.

\vskip 1cm
\noindent
{\bf \Large References and Notes}

\bigskip
\noindent
1. \ \ http://www.vla.nrao.edu

\smallskip
\noindent
2. \  Heeschen, D.S., The Very Large Array, in Telescopes for the 1980s,
G. Burbidge and A. Hewitt,
editors, Annual~Reviews~Inc., 1981, pp. 1-61

\smallskip
\noindent
3. \ \ T.J.~{Pearson}, A.C.S~{Readhead}, 1984, {\it Ann. Rev. Astron. Astrophys. \/} {\bf 22}, 97

\smallskip
\noindent
4. \ \ Kaastra J.S. \& Mewe R.
1993, {\it Astron. Astrophys. Suppl. \/} {\bf 97}, 443

\smallskip
\noindent
5. \ \ Liedahl D.A., Osterheld A.L.
\& Goldstein W.H. 1995, {\it Astrophys. J. \/} {\bf 438}, L115

\smallskip
\noindent
6. \ \ Anders E., Grevesse N.,
1989, {\it Geochimica et Cosmochimica Acta \/} {\bf 53}, 197

\smallskip
\noindent
7. \ \ Grevesse N., Sauval A.J., 1998,
{\it Space Science Reviews \/}{\bf 85}, 161

\smallskip
\noindent
8. \ \ Wilms J., Allen A., McCray R., 2000,
{\it Astrophys. J. \/} {\bf 542}, 914

\smallskip
\noindent
9. \ \ Balucinska-Church
M. \& McCammon D. 1992, {\it Astrophys. J. \/} {\bf 400}, 699

\smallskip
\noindent
10. \ \ Lumb D.H., Warwick R.S., Page M., \& De Luca A. 2002, A\&A {\bf 389}, 93

\smallskip
\noindent
11. \ \ Durret F., Lima Neto G.B., Forman W., 2005, A\&A {\bf 432}, 809

\clearpage
\begin{figure}
\begin{center}
\epsfig{width=7.8cm,file=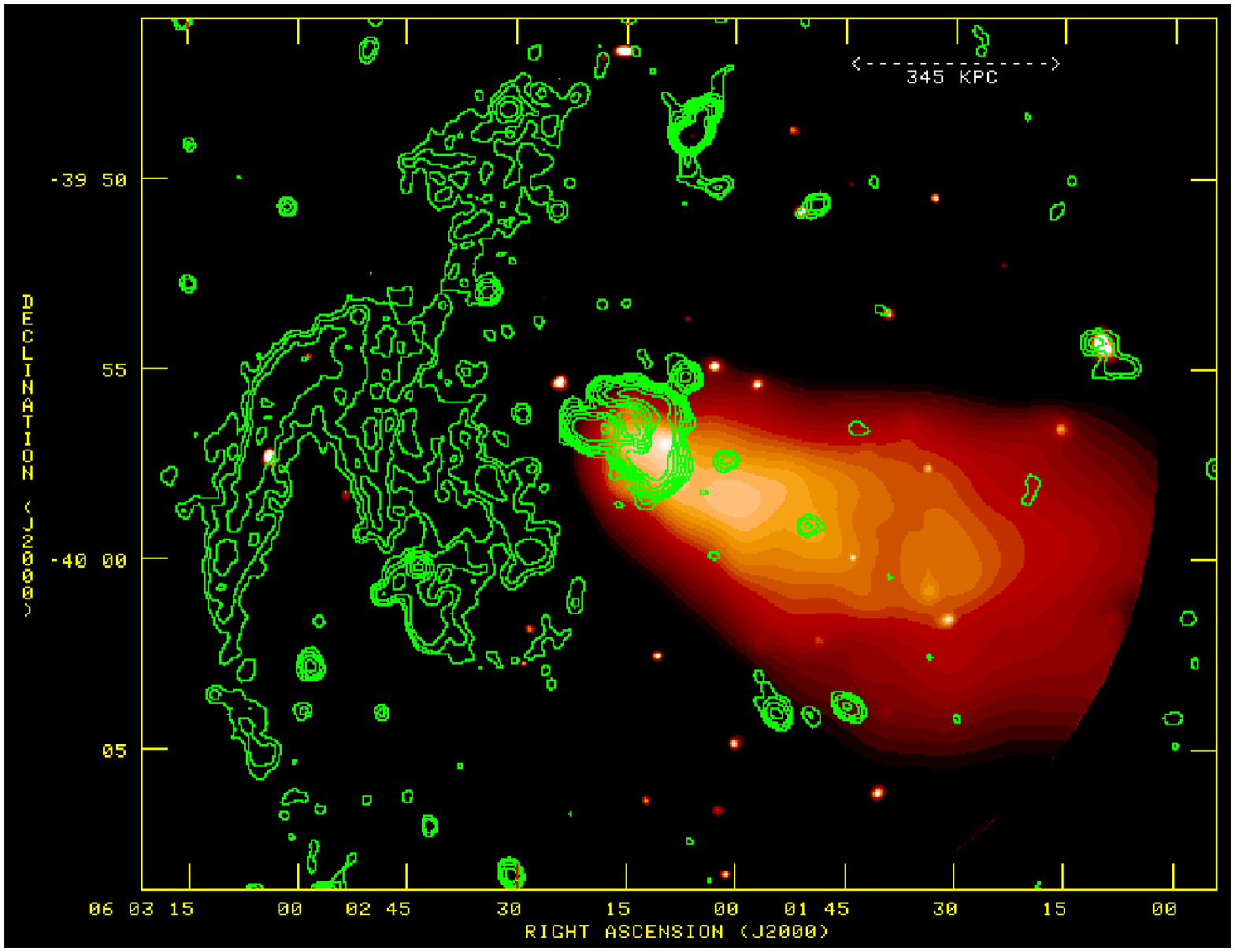,angle=-90}
\end{center} 
{\bf Figure~S1(a):} {\footnotesize Image details near the strong 
radio source MRC~0600-399 in \clust cluster, which
is positioned near the peak of ``cometary'' thermal bremsstrahlung 
X-ray emission (main text). 
The 1.4 GHz radio emission observed by Very~Large~Array, at 20 arcsec beam resolution
(contours), are shown superposed on the XMM-Newton X-ray map (0.3--8.0 keV band). 
Contour values are: (0.12, 0.24, 0.48 and 1 mJy/beam).
We  can see the X-ray
emission associated with the radio galaxy core and a pair of radio jets strongly bent
towards north north-east, away from the elongated X-ray emission. Also visible is the 
large scale diffuse radio emission further to the east.}
\end{figure}
\begin{figure}
\begin{center}
\includegraphics[width=5cm]{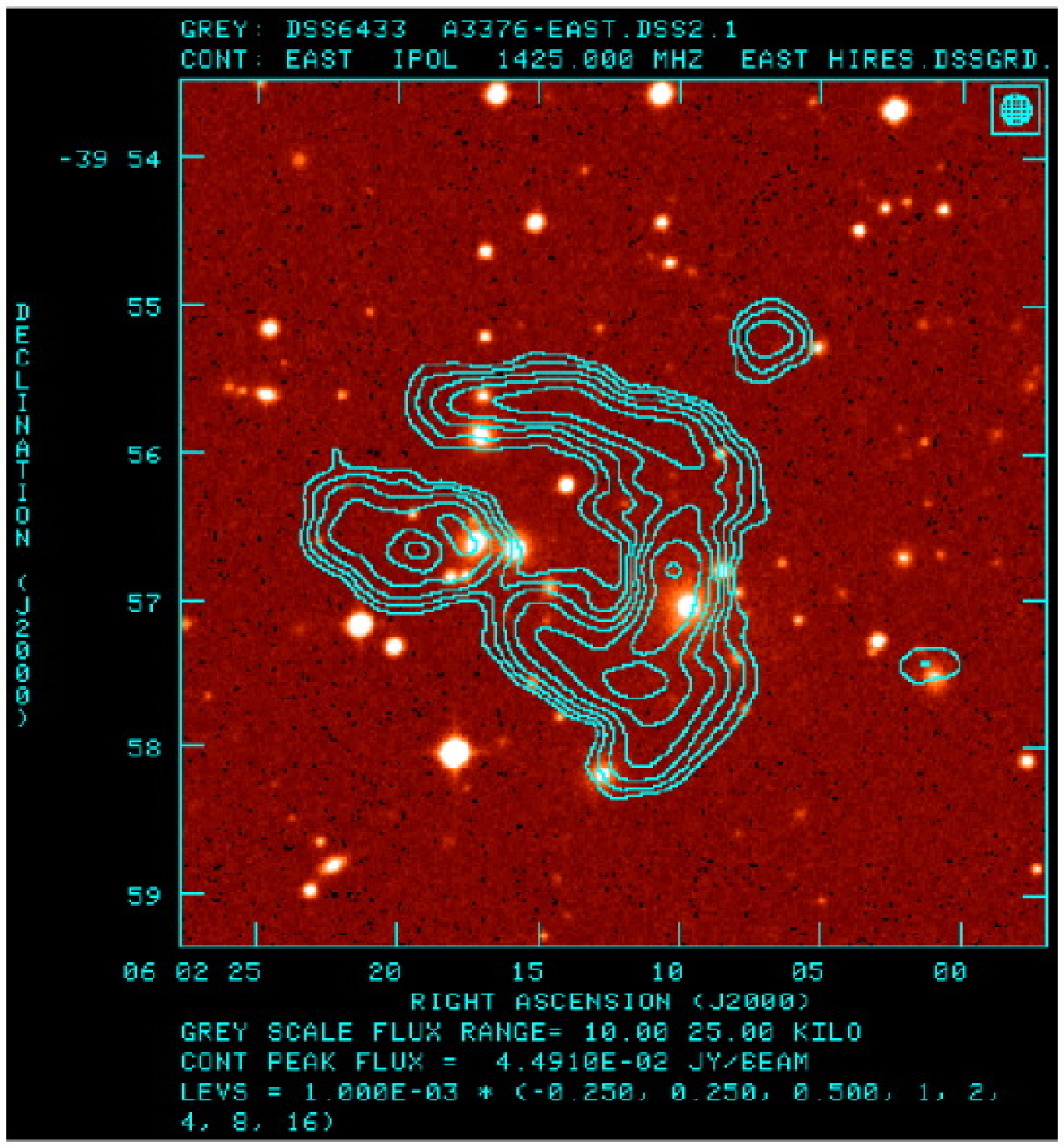}
\end{center}
{\bf Figure~S1(b):} {\footnotesize Image showing a 
superposition of radio emission contours  
(at 1.4 GHz, observed with Very~Large~Array, at 12 arcsec beam resolution), 
and the r-band
optical image from the Digitized Sky Survey. The 2nd brightest  cluster member
galaxy, associated with  the bent-jet radio galaxy MRC~0600-399 can be seen 
close to its
radio core. Another radio source, possibly originating from an elliptical galaxy
to the north east of MRC~0600-399 can also be seen. The image covers
$\approx 5 \times 5$ arcmin$^2$ ($300 \times 300$ kpc$^2$) region. 
Contour values are : (0.25, 0.5, 1, 2, 4, 8 and 16 mJy/beam)} 
\end{figure}

\clearpage
\begin{figure}
\begin{center}
\epsfig{width=15cm,file=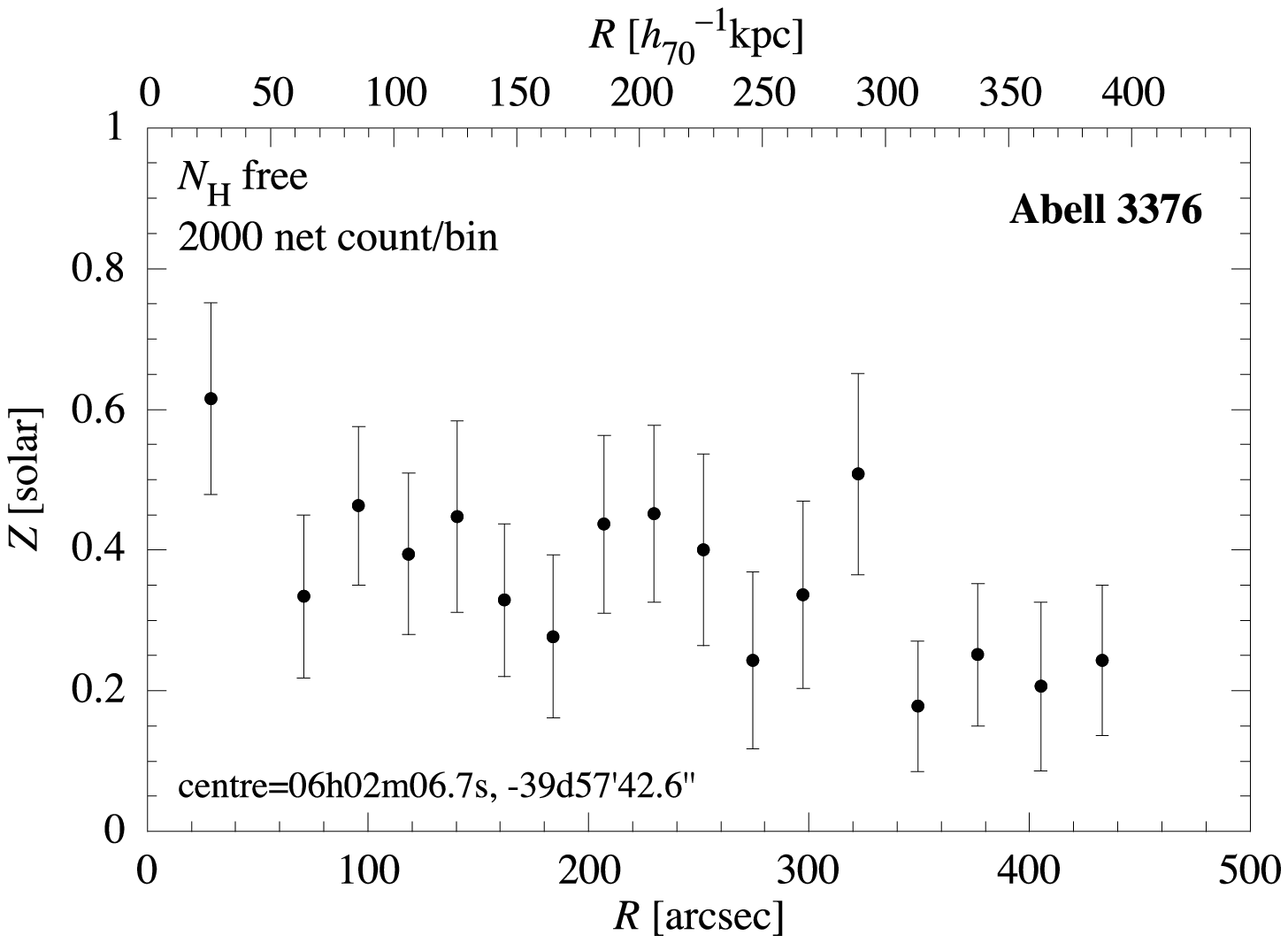}
\end{center}
{\bf Figure~S2:} Azimuthally averaged, radial profile of the intra-cluster 
medium metallicity (Z), here given in solar
units, obtained from analysis of XMM-Newton data. The center of the 
profile is located on the X-ray peak position 
R.A. $
06^h~02^m~6.7^s$, and Dec. $ -39^\circ 57^\prime
42.6^{\prime\prime}$.
The metal abundance is mainly based on detection of
Fe~XXV K-$\alpha$ complex at $\sim
6.7$~keV (restframe energy). The neutral hydrogen coloumn density is taken
as free parameter for fitting and each bin contains $> 2000$ counts. 
There appears to be no  strong metallicity gradient, as is the case of
virialized clusters, although a weak trend can be seen. 
Some dynamically virialized  
clusters can have metallicity as high as the $1
Z_{solar}$ in the centre, see, e.g., ({\it 10}). The absence of a clear
metallicity gradient is a sign of mixing, possibly caused by  a 
recent merging event of subclusters with their varied individual 
metal abumdances. 
\end{figure}

\clearpage
\begin{figure}
\begin{center}
\epsfig{width=15.5cm,file=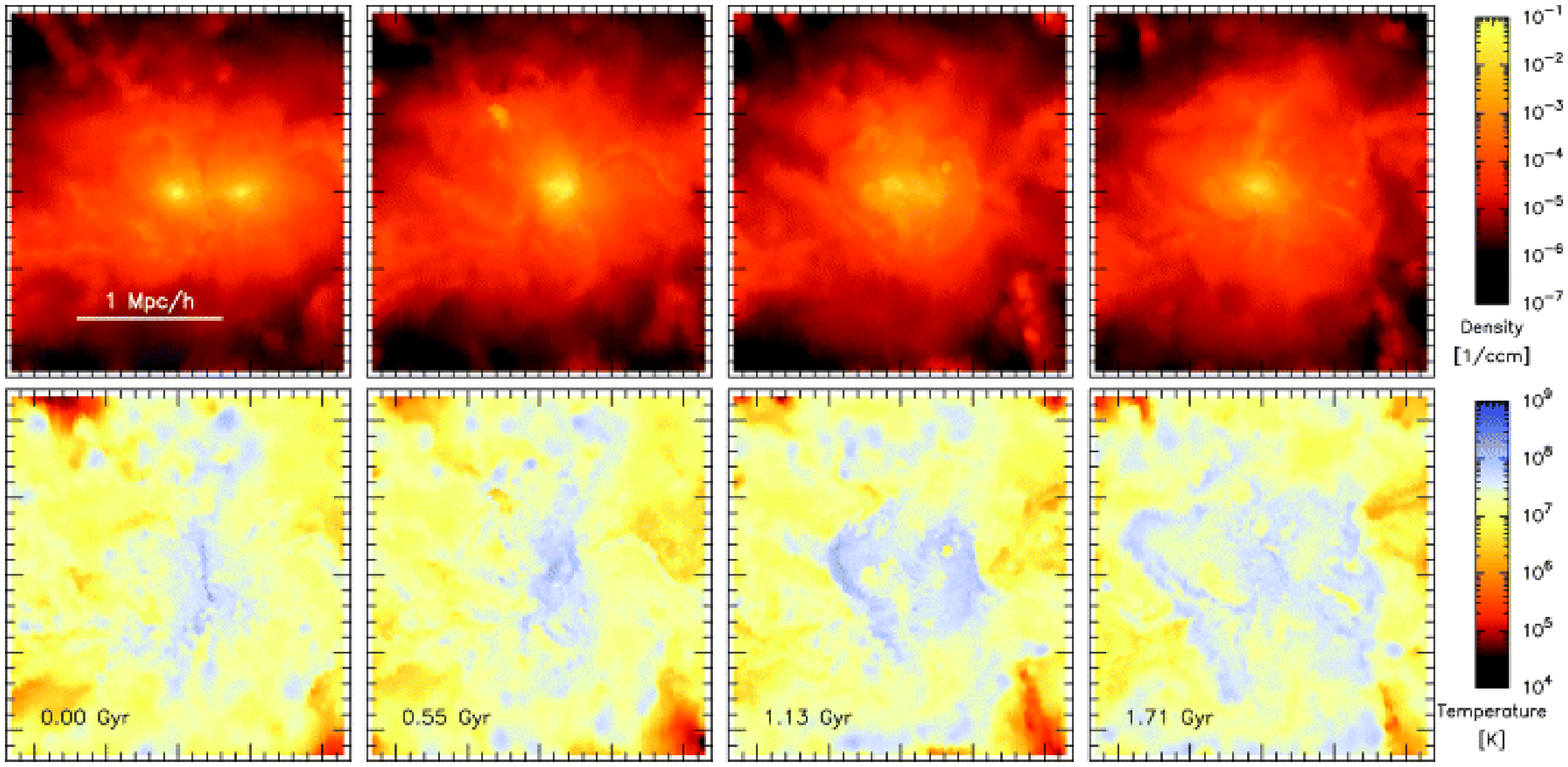}
\end{center}
{\bf Figure~S3 (a):} Results from a hydrodynamical simulation (with gas dynamics and
magnetic fields `frozen' into gas) 
of merger
between two clusters of similar masses, $1.7 \times 10^{13} M_{\odot}$ 
and $1.6 \times 10^{13} M_{\odot}$ (at redshift $z\approx0.6$). The collision between
two masses is virtually head-on with a relative speed of 2000 km/s. We see
the density (upper
panels) and temperature (lower panels) distribution on a slice along a plane which contains
the both centres of mass of the initial two clusters. The scale bar is comoving length
in $h^{-1}$ Mpc. The first snapshot is taken at redshift $z =0.66$ at $t = 0$, and 
later snapshots, with snapshot time indicated in Gyr (1 Gyr = $10^{9}$ yr),
follow the center of mass of the cluster. The density  is given in proper
mass density divided by proton mass, and the gas temperature is in Kelvin
degrees. In the temperature maps, we can clearly see  a pair of 
very hot ($\sim 10^{8}$ K),
merger generated shock
fronts propagating outwards from the cluster center, reaching $\approx$ 1 Mpc
distance at the outskirts of the final collapsed cluster. The structure of these
`ring-like' shocks is very similar to giant radio arcs discovered in \clust 
cluster. Many other filamentary shock features can also be seen, similar to what we
see towards west of the eastern radio structure found in cluster 
\clust (Fig.~1, and main text).
(Simulation figure: courtsey of, Matthias Hoeft and Marcus Bruggen. For more details 
see the original paper:
M. Hoeft, M. Bruggen and G. Yepes, {\it Mon. Not. R. Astron. Soc. \/}
{\bf 347}, 389 (2004).
\end{figure} 
\clearpage
\begin{figure}
\begin{center}
\epsfig{width=15.5cm,file=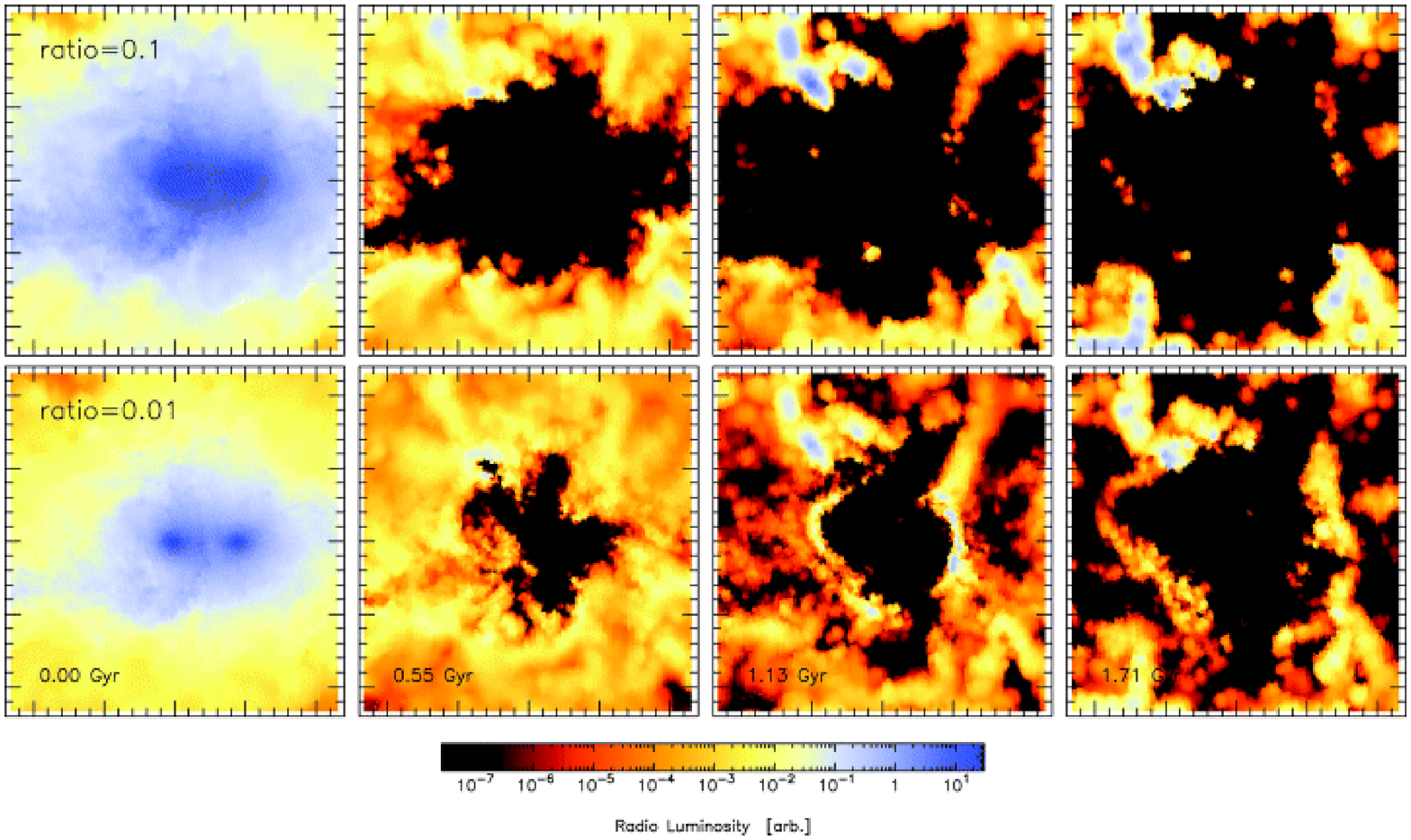}
\end{center}
{\bf Figure~S3 (b):} Results from a hydrodynamical simulation of merger shocks and
creation of large scale radio relics.
This figure shows the radio luminosity probability at a
frequency of 100 MHz, per unit mass of radio plasma. Other parameters are the
same as in Fig.~S3(a). Thus, this figure shows not the actual radio luminosity, but the
scaled probability of radio emission in arbitrary units (see the color bar. Black
or red color show low probability, while blue and yellow trace regions of higher
radio emission). The upper panel is for higher value ($P_{B}/P_{gas}= 0.1$) of the
ratio of magnetic pressure
($P_{B}$) to gas pressure ($P_{gas}$). The lower panel is for lower value (0.01) of
the same ratio. The most remarkable structure is a prominant ring-like feature with 
diameter of about 1 Mpc,  1.13 Gyr after the shock fronts were generated from the
center due to a merger. This radio structure corresponds to the outgoing shock waves
seen in Fig.~S3(a). One can clearly see the flaring of a radio plasma due to 
in situ shock acceleration of electrons and compression of radio plasma, 
even after 1 Gyr past of merger. Higher
 magnetic fields (upper panels) result in too fast an ageing of the plasma (due
 to synchrotron emission losses), such that
 the shock waves can not `revive' the plasma. Noticeably, the  center 
 of cluster is virtually
 void of any 'radio-halo' type synchrotron emission. The  lack of 
 a `radio-halo' emission from
 cluster center and the simulated structure of these
 `ring-like' shocks, both are  remarkably similar to the peripheral
 giant arc-like radio structures 
 we discovered in \clust
 cluster (Fig~.1, and  main text).
 (Simulation figure: courtsey of, Matthias Hoeft and 
 Marcus Bruggen. For more details
 see the original paper:
 M. Hoeft, M. Bruggen and G. Yepes, {\it Mon. Not. R. Astron. Soc. \/}
 {\bf 347}, 389 (2004).
\end{figure}

\end{document}